\documentclass[11pt]{article}
\usepackage{xspace,amssymb,amsmath,helvet,graphicx,natbib,url}
\begin{document}
% useful commands
\newcommand{\dee}{\,\mbox{d}}
\newcommand{\naive}{na\"{\i}ve }
\newcommand{\eg}{e.g.\xspace}
\newcommand{\ie}{i.e.\xspace}
\newcommand{\pdf}{pdf.\xspace}
\newcommand{\etc}{etc.\@\xspace}
\newcommand{\PhD}{Ph.D.\xspace}
\newcommand{\MSc}{M.Sc.\xspace}
\newcommand{\BA}{B.A.\xspace}
\newcommand{\MA}{M.A.\xspace}
\newcommand{\role}{r\^{o}le}
\newcommand{\signoff}{\hspace*{\fill} Rose Baker \today}
% entry environment
\newenvironment{entry}[1]%
{\begin{list}{}{\renewcommand{\makelabel}[1]{\textsf{##1:}\hfil}%
\settowidth{\labelwidth}{\textsf{#1:}}%
\setlength{\leftmargin}{\labelwidth}
\addtolength{\leftmargin}{\labelsep}
\setlength{\itemindent}{0pt}
}}%
{\end{list}}
\title{A flexible and computationally tractable discrete distribution derived from a stationary renewal process}
\author{Rose Baker\\School of Business\\University of Salford, UK}
%\email{rose.baker@cantab.net}
\maketitle
\begin{abstract}
A class of discrete distributions can be derived from  stationary renewal processes. They have the useful property that the mean is a simple function of
the model parameters. Thus regressions of the distribution mean on covariates can be carried out and marginal effects of covariates calculated.
Probabilities can be easily computed in closed form for only two such distributions, when the  event interarrival times in the renewal process
follow either a gamma or an inverse Gaussian distribution. The
gamma-based distribution has more attractive properties and is described and fitted to data. The inverse-Gaussian based distribution is also briefly discussed.

\end{abstract}
\section*{Keywords}
Equilibrium renewal process; gamma function; discrete distribution; marginal effect; closed-form solution; inverse Gaussian distribution.
\section{Introduction}
Discrete distributions are used when modelling count data, and the dependence of counts on covariates.
There is a very wide range of application areas, \eg  life sciences, economics, maintenance and reliability.

The Poisson distribution is the best-known discrete distribution. However, count data often show overdispersion, or, more rarely, underdispersion, and the probability of occurrence of zero events 
often differs from what the Poisson distribution would predict. Very many 2-parameter discrete distributions exist (\eg \citet{john}), derivable in many ways;
here the focus is on distributions that can be derived from renewal processes (RPs). \footnote{For completeness, 
a renewal process is defined here. Let $S_n=\sum_{i=1}^n X_i$ be a sum of $n$ i.i.d. positive random variables. The counting process $N(t)=\text{max}(n: S_n \le t)$, the number of events that have occurred 
by time $t$, is a renewal process.}

Most count data (but not all) are derived from processes occurring in time, such as number of doctor visits over a period, number of children born, \etc, so that an RP
may be an approximation to the underlying process that generated the observed counts. The Poisson and negative binomial distributions can be derived from Poisson processes, which are RPs, and so fall in this class.

Among several desirable properties for a discrete distribution an important one is that the mean $\eta=\text{E}(N)$ should be simply expressible in terms of the model parameters. A major interest is the dependence of the distribution
on a vector of covariates ${\bf x}$, and this is usually expressed as $\eta=\eta_0\exp({\boldsymbol \beta}^T{\bf x})$, where ${\boldsymbol \beta}$ is a vector of coefficients.
Modelling the distribution mean as a function of covariates gives an easily-interpretable model, from which economists and others can make simple calculations. Thus, in the example
of completed birth rate given later, it becomes trivial to ask how many more or how many fewer children should be born in the population if 50\% of women had received a university education.

For distributions derivable from Poisson processes, the mean is easily calculable. However, in general this is not so, and the best that can be done theoretically is to present
an asymptotic form for the expected number of events by time $t$ as $\text{E}(N(t)) \simeq \frac{t}{\mu}+\frac{\sigma^2-\mu^2}{2\mu^2}$, where $\mu$ and $\sigma^2$ are the mean and variance of the interarrival time,
\eg \citet{cox}. In practice one would need to compute the mean exactly for two different values of the covariate to read off the marginal effect of the covariate. This can be done, but is a further step of analysis.
Also, it is desirable for simplicity of interpretation to model the mean of the discrete distribution as a function of covariates, not the mean of the interarrival process.

To derive a discrete model with a simple formula for the mean,
it is necessary to consider a stationary (equilibrium) RP, an ERP. In this case $\text{E}(N(t))=t/\mu$. 
%One notices that a Poisson process is also a stationary RP,
%which explains why the Poisson distribution and those derived from it by mixture have means $\eta$ of this form.
One can conveniently name discrete distributions derived from renewal processes as RP-X, ERP-X, where $X$ is the name of the interarrival time distribution.

Quite a lot of work has been done to compute count probabilities arising from RPs.
The connection between renewal processes and discrete distributions is discussed in \citet{cox},
who describes the distributions arising from the Erlang RP, and the negative binomial as a mixture of Poisson RPs with gamma-distributed stopping times.
The gamma RP case has been developed by \citet{winkl}.
Computations for the Weibull RP were considered by \citet{shane} following \citet{lom}, and these models have also been used in sport analytics \citep{bosh}.
The Mittag-Leffler distribution has also been used \citep{mittag}.
An excellent summary is given in \citet{Jose2013Gumbel}.

Stationary RPs have hardly been explored, but the R language Countr package \citep{Countr,bakelast}
allows modified RPs, of which a stationary RP is a special case. Hence models based on these processes are becoming available to the user.

Here however, the aim is to explore the two distributions where it turns out that the probabilities can conveniently be derived in closed-form, \ie in terms of special functions.
The ERP with gamma-distributed interarrival times gives rise to a discrete distribution, called here the ERP-$\gamma$ distribution.
The corresponding distribution derived from an ordinary RP has been used by \citet{winkl}, and allows for both under and overdispersion. He comments
\citep{winkelmann2013econometric} on the `small catch' that the mean is not calculable, and this paper addresses this problem.

Since \citet{cox} has mentioned the ERP-$\gamma$ distribution in the content of the distribution of the number of renewals, it cannot be claimed as a new discrete distribution.
Rather, the original contribution here is to propose this distribution as a useful distribution for count regression, and to show how the necessary computations can be done with it.

The discrete distribution based on the inverse Gaussian (IG) distribution (the ERP-IG distribution) has not been looked at before. Although it is easier to compute with than the ERP-$\gamma$
distribution, it does not contain the Poisson distribution as a special case, and it did not fit the example data even as well as the Poisson. However, it may prove useful in some contexts and so is briefly described in appendix A.

The next section introduces some notation and discusses the distribution used by Winkelmann. Next, the probabilities for the ERP-$\gamma$ distribution are derived, and its properties given. 
Some extensions of the ERP-$\gamma$ and RP-$\gamma$ distributions are discussed.
The new distribution is fitted to a well-used dataset of fertility (number of children as a function of mother's age, \etc) to demonstrate its feasibility.

\section{The ERP-$\gamma$ distribution}
\subsection{Definitions and Notation}
To introduce some notation, the gamma distribution probability density function (pdf) is
$f(t;\alpha,\beta)=\alpha (\alpha t)^{\beta-1}\exp(-\alpha t)/\Gamma(\beta)$, 
with (cumulative) distribution function (cdf) $F(t)$ and mean $\mu=\beta/\alpha$. 

To be amenable to computation, ERP and RP distributions must possess the additive property, that a sum of i.i.d. random variables from the distribution belongs to the same family of distributions.
Both the gamma and inverse Gaussian distributions possess this property. This is sometimes called the reproductive property, which term is commonly used in a more general sense,
\ie that sums of random variables from the distribution family, but with different parameters, belong to the same distribution family. Both the gamma and IG distributions also have this property,
\eg a sum of gamma r..v.s with different $\beta$ parameters is gamma. This property allows greater flexibility in constructing RP distributions, but cannot be invoked for ERP
distributions without the process ceasing to be an ERP, and so losing the simple formula for the mean. Finally there is the still more general  property of divisibility, which means that a random variate can be decomposed into
two or more others, not necessarily from the same distribution, and which is not needed here.

There may be infinitely many survival distributions with the additive property, \eg the class of exponential dispersion models \citep{jorg}. The Tweedie distributions belong to this class \citep{jorg},
and both the gamma and inverse Gaussian distributions are Tweedie distributions. There do not seem to be any others that are computationally tractable; \eg, the compound Poisson-gamma
distribution is also a Tweedie distribution, but the pdf must be expressed as a Bessel function.

By the additive property of the gamma distribution, the sum of $n$ gamma random variables
has pdf $f^{(n)}(t)=f(t;\alpha,n\beta)$. Let the corresponding cdf be $F^{(n)}(t)$, also written as the incomplete (regularised) gamma function
$\gamma(x)=\int_0^x u^{n\beta-1}\exp(-u)\dee u/\Gamma(n\beta)$. Successive events at times $X_1, X_1+X_2 \cdots$ form an RP
with $N(t)$ events having occurred by time $t$. The probability of $n$ events is $\text{Prob}(N(t)=n)\equiv P_n(t)=F^{(n)}(t)-F^{(n+1)}(t)$ for $n > 0$, and $P_0(t)=1-F^{(1)}(t)$.
This is Winkelmann's distribution. 

The corresponding pdf for the distribution proposed here is $g(t;\alpha,\beta)$, with cdf $G^{(n)}(t)$ and count probabilities $Q_n(t)$.
The arbitrary stopping time $t$ is retained throughout, but without loss of generality one can set $t=1$.

Note that the results here have been checked by simulating count probabilities from the distributions and comparing with the formulae derived.
The fortran prototype programs, which use NAG (Numerical Analysis Group) library routines, are available online.

\subsection{Derivation of the ERP-$\gamma$ probability mass function (pmf)}
The derivation of the pmf has two steps: first, obtaining the probabilities $Q_n(t)$ as an integral, then showing that the integral can be evaluated in terms of the incomplete gamma function.
In an equilibrium RP, the time to first event has pdf $S(t)/\mu$, where $S$ is the survival function, \ie $S(t)=1-F(t)$, and so the cdf for $n$ events is
\begin{equation}G^{(n)}(t)= \mu^{-1}\int_0^t\int_0^u S(w)f^{(n-1)}(u-w)\dee w\dee u.\label{eq:gn}\end{equation}
This equation means that the first event occurs at time $w$, and $n-1$ events then occur by time $u$, so at least $n$ events have occurred by time $t$.
We can also write for an RP
\[P_n(t)=\int_0^t f^{(n)}(u)S(t-u)\dee u,\]
which means that exactly $n$ events have occurred by time $t$ when $n$ events have occurred by some time $u$ and no further events then occur.
The similarity of this equation to (\ref{eq:gn}) can be exploited to obtain
\begin{equation}G^{(n)}(t)=\mu^{-1}\int_0^t P_{n-1}(u)\dee u\label{eq:gn2}\end{equation}
for $n > 0$; of course, $G^{(0)}(t)=1$.

Hence the probabilities $Q_N(t)$ can be written as:
\[Q_0(t)=1-\mu^{-1}\int_0^t P_0(u)\dee u,\]
\[Q_1(t)=\mu^{-1}\int_0^t (1-P_1(u))\dee u,\]
and for $n > 1$
\begin{equation}Q_n(t)=\mu^{-1}\int_0^t\{F^{(n-1)}(u)-2F^{(n)}(u)+F^{(n+1)}(u)\}\dee u.\label{eq:myq}\end{equation}
Equation (\ref{eq:myq}) is true for any RP, and is found in \citet{cox}, where it is derived using Laplace transforms,
rather than by the probabilistic argument used here. From here on, the derivation is specific to the ERP-$\gamma$ distribution.

The second step evaluates integrals such as 
\begin{equation}I_n=\int_0^tF^{(n)}(u)\dee u=\int_0^t\int_0^u f^{(n)}(w)\dee w\dee u,\label{eq:in0}\end{equation}
where $n > 0$.
Exchanging the order of integration or integrating by parts, we have that
\[I_n=\int_0^t (t-w)f^{(n)}(w)\dee w=t\gamma(\alpha t; n\beta)-(n\beta/\alpha)\gamma(\alpha t; n\beta+1).\]
The last term can be simplified by integrating by parts, to finally obtain
\begin{equation}I_n=(t-n\beta/\alpha)\gamma(\alpha t; n\beta)+\alpha^{-1}(\alpha t)^{n\beta}\exp(-\alpha t)/\Gamma(n\beta).\label{eq:in}\end{equation}

The probabilities are now
\[Q_0(t)=1-t/\mu+I_1(t)/\mu,\]
\[Q_1(t)=t/\mu+\mu^{-1}\{I_2(t)-2I_1(t)\},\]
and for $n > 1$
\[Q_n=\mu^{-1}\{I_{n-1}(t)-2I_n(t)+I_{n+1}(t)\}.\]

The cdf is needed when there is censoring, so that for example large counts are recorded as being greater than some count $M$.
This is
\[G_1(t)=t/\mu-I_1(t)/\mu,\]
\[G_{n>1}=\mu^{-1}\{I_{n-1}(t)-I_n(t)\}.\]

\subsection{Properties}
\begin{figure}[h]
\centering
\makebox{\includegraphics{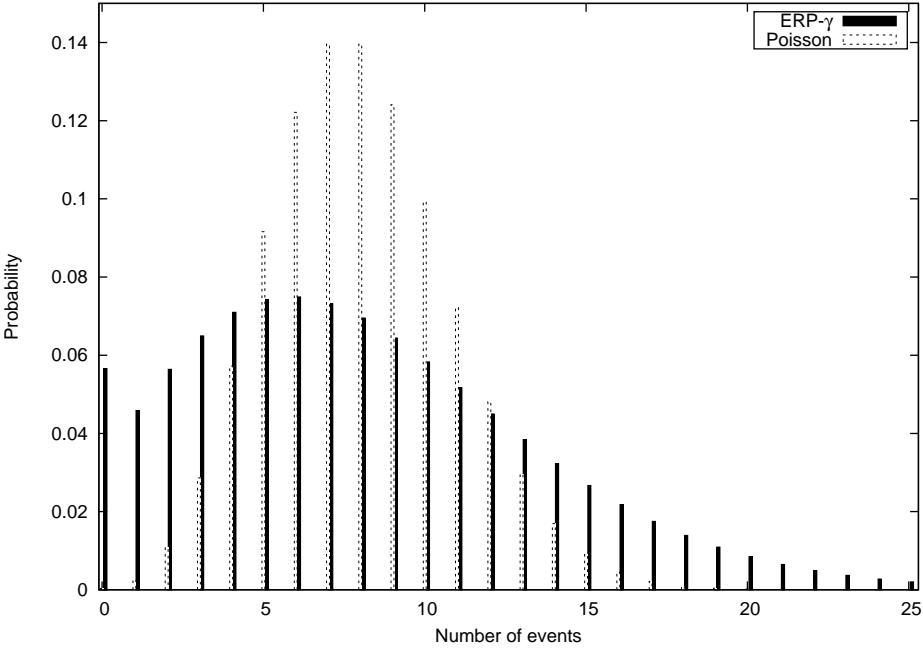}}
\caption{\label{figb}Overdispersed ERP-$\gamma$ distribution with the Poisson; mean 8, $\alpha=2, \beta=0.25$.}
\end{figure}

\begin{figure}[h]
\centering
\makebox{\includegraphics{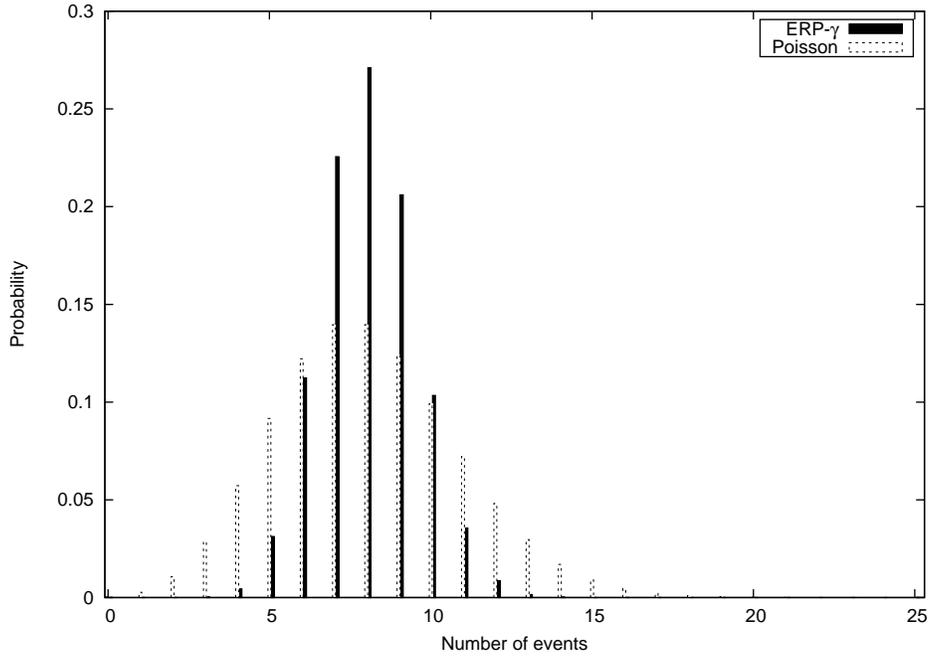}}
\caption{\label{figb2}Underdispersed ERP-$\gamma$ distribution with the Poisson; mean 8,  $\alpha=32, \beta=4$.}
\end{figure}

From the formulae for the probabilities $Q$ it is trivial to verify that $\sum_{i=0}^\infty Q_i=1$ and that $\text{E}(N(t))=t/\mu$.
The formula for the variance simplifies to $\text{var}N(t)=(2/\mu)\sum_{i=1}^\infty I_i(t)+(t/\mu)(1-t/\mu)$. If $\beta > 1, \text{var}(N(t)) < t/\mu$ so the distribution is underdispersed, 
and if $\beta < 1$ it is overdispersed. From the general formula for the asymptotic variance of an equilibrium RP given in \citet{coxmill} we have that
\[\text{var}N(t)=\frac{\alpha t}{\beta^2}+\frac{1}{6}+\frac{1}{2\beta^2}-\frac{2}{3\beta^{1/2}}+o(1),\]
which works well (used as an exact formula) when $\alpha t \gg 1$. The distribution is asymptotically normal, but can be overdispersed or underdispersed,
as shown in figures \ref{figb} and \ref{figb2}. Figure \ref{figb} shows a peak at zero. Intuitively, this arises because the first random event has a distribution with higher mean than the others, if $\beta < 1$.
The peak at zero arises when the distribution is very overdispersed, but it is absent for more modest overdispersion. When the mean is low enough, the overdispersed distribution is J-shaped.

Random numbers can be generated for the RP-$\gamma$ distribution by generating random numbers $X_1, X_2\cdots$ from the gamma distribution,
and counting how many numbers it takes for the sum $X_1+X_2+\cdots$ to exceed $t$ (the discrete random number is 1 less than this). For the ERP-$\gamma$ distribution
the first random number comes from a different distribution. However, as ever the solution is implicit in \citet{cox}. He gives a derivation of the time to first event
by considering the length biased pdf $xf(x)/\mu$ \citep[section 5.4]{cox}. Following this argument, one can generate the time to first renewal by generating a random number $Y$ from the 
gamma $(\alpha,\beta+1)$ distribution, and then taking the time to first renewal as $UY$, where $U$ is a random number from the $[0,1]$  uniform distribution.

The ERP-$\gamma$ distribution does have some causal basis. Sometimes one starts collecting data when a counting process akin to a renewal process is already underway, \eg one starts 
counting failures of equipment that is already in use. In this case, the ERP-$\gamma$ distribution is a flexible model of what is happening. In many other cases,
the connection to a renewal process will be vaguer, and sometimes, as when for example counting bacteria on a microscope slide where the process does not occur in time at all,
the distribution is simply a mathematically and computationally convenient choice.

\subsection{Other related distributions}
It is worth mentioning that Winkelmann's RP-$\gamma$ distribution can be easily generalised into a simple hurdle model, by exploiting the reproductive property of the gamma distribution.
The time to the initial event can have shape parameter $\beta+\delta$, where $\delta > -\beta$.
Then the probability of $n$ events is $P_n(t)=\gamma(\alpha t; n\beta+\delta)-\gamma(\alpha t; (n+1)\beta+\delta)$, $P_0=1-\gamma(\alpha t;\beta+\delta)$.
This allows the probability of zero events to be varied, and a test of whether a hurdle is present or not to be done.

One can of course in general make the $m$th interarrival time different from the others. Then
\[P_n(t)=\gamma(\alpha t; n\beta+\theta(n-m)\delta)-\gamma(\alpha t; (n+1)\beta+\theta(n+1-m)\delta),\]
\[P_0=1-\gamma(\alpha t;\beta+\theta(1-m)\delta),\]
where $\theta$ is the discrete Heaviside step function; $\theta(n)=0$ if $n < 0$, else $\theta(n)=1$.

In the case of fertility, one might suppose that many parents decide to stop after having two children,
so the third interarrival time could be increased by increasing $\beta$ to $\beta+\delta$. 
Precisely this model gives the best fit to the fertility data used in the example in the next section.

\section{Example}
This is the completed fertility dataset from the second (1985) wave of the German Socio-Economic Panel, described in \citet{winkl}. It contains number of children (0-11)
and 10 demographic covariates for 1243 women. The count distribution is slightly underdispersed, and becomes more so after regressing on the covariates.
Six distributions were fitted: the Poisson, Winkelmann's RP-$\gamma$ distribution, the ERP-$\gamma$ distribution, a mixture of ERP-$\gamma$ distributions with different $\beta$ values, ditto with different $\alpha$ values,
and the RP-$\gamma$(3) distribution. In addition, \citet{shane} has fitted a distribution with Weibull interarrival times and the heterogeneous Weibull distribution.

\begin{figure}[h]
\centering
\makebox{\includegraphics{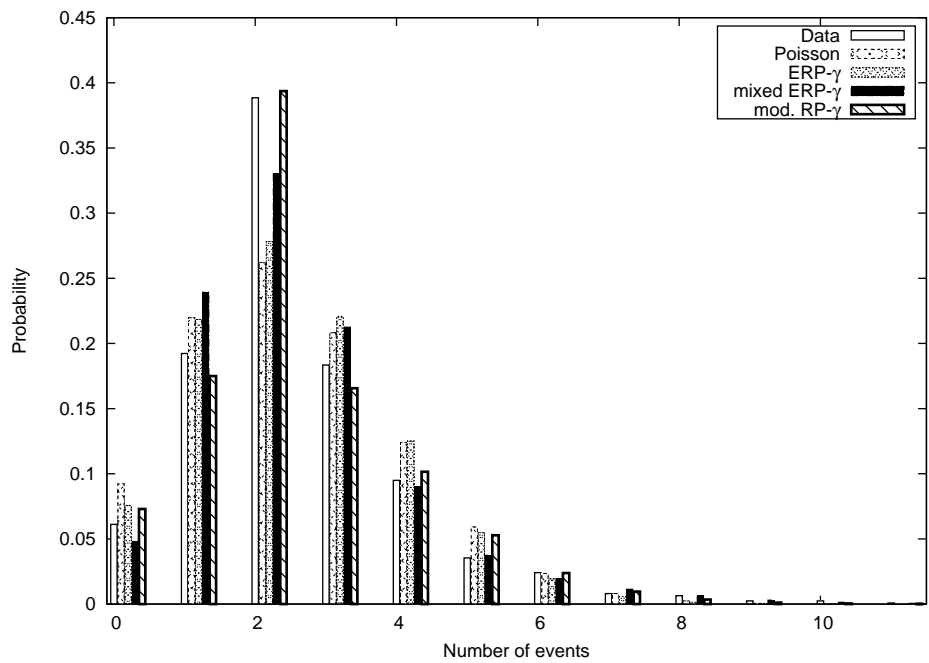}}
\caption{\label{figa}Fertility data at left and towards right Poisson, ERP-$\gamma$, mixed ERP-$\gamma$, and modified RP-$\gamma$ fits.}
\end{figure}

First, omitting all covariates, table \ref{tab1} shows the fitted parameters and minus the log-likelihood values, and figure \ref{figa} shows the data and some of the fitted distributions.
It can be seen that the ERP-$\gamma$ fits slightly better than the RP-$\gamma$, so using an equilibrium RP has not worsened the fit.
The mixture of ERP-$\gamma$ distributions obtained by using two values of $\beta$ requires two additional parameters, and gives a much better fit; almost the same can be achieved by varying $\alpha$.
However, the best fit results with the RP-$\gamma$(3) distribution, with $\delta=0.66$ added. This probably best reflects the underlying reality, that probably many couples decide that two children is enough.
The `hazard' of producing the third child is reduced.

For completeness, table \ref{tab2} shows the fit results obtained with covariates, using the RP-$\gamma$ and ERP-$\gamma$ distributions. They are very similar,
but in general, the ERP-$\gamma$ coefficients are slightly larger, as are the standard errors. 
The ERP-$\gamma$ results have the merit of being more easily interpretable. The estimated mean ${\text{E}(N(1))}=2.314$, and so the estimated marginal effects are just the
coefficient values multiplied by this number.

The marginal effect $\partial \text{E}(N|{\bf x})/\partial x_j=\beta_j\text{E}(N|{\bf x})$ is also trivial to calculate given the model fit, and the standard error
of the marginal effect can be found by applying the delta-method and using the estimated covariance matrix on the fitted model parameters. 

\section{Conclusions}
A new class of discrete distributions based on an equilibrium renewal process has been introduced. A member of this class where the probabilities can be written in closed-form
has been derived, its properties discussed, and fitted to data. This is a flexible distribution that generalizes the Poisson, 
can model under or over-dispersion, and which allows marginal effects to be computed.
Computation of probabilities requires only the incomplete gamma function, available on just about every computing platform.

Hurdle models are widely used to model an excess of zero counts. Introducing a hurdle directly would mean that the RP was no longer an equilibrium RP, and so the simple expression $t/\mu$ for the mean
and the ability to easily compute marginal effects would go. However, at the cost of introducing two extra parameters, one can model a hurdle as a mixture of two ERP-$\gamma$
distributions with different values of the shape parameter $\beta$.

Extensions to Winkelmann's RP-$\gamma$ distribution have also been introduced. This distribution is simpler to compute with than the ERP-$\gamma$ distribution,
but does not allow easy computation of the mean and hence of marginal effects. Having sacrificed this property, however, one can introduce hurdles at any point, \eg in the example, a hurdle after the birth of two children.
This requires only one additional parameter. 

The inverse Gaussian distribution has also been discussed. As it does not contain the Poisson distribution as a special case,
it does not currently look as attractive as the ERP-$\gamma$ distribution, but it may yet find applications.
Compared to a Poisson distribution of the same mean and variance it has more probability at zero, a peak shifted slightly to the right, and a shorter tail.

Further work could proceed on two fronts, the first being the search for more ERP distributions with tractable computational properties.
Also extensions with additional parameters that are easy to compute would add more flexibility. This could be done by allowing the termination time $t$ to have a distribution,
or using a 3-parameter distribution for interarrival times.
The second front is experience with other distributions such as the ERP-Weibull,
for which finding probabilities requires more extensive computation, but is still quite feasible. 
\bibliographystyle{apalike}
\bibliography{nrefs}
\section*{Appendix A: Inverse-Gaussian (IG) interarrival times}
The basic facts about the inverse Gaussian distribution are taken from \citet{kotz} and \cite{chik}. 
The distribution arises as the time to first passage through a barrier for a particle undergoing a Gaussian random walk.
The pdf is
\[f(x)=(\frac{\lambda}{2\pi x^3})^{1/2}\exp(-\lambda (x-\mu)^2/2\mu^2 x), \]
with mean $\mu$, variance $\mu^3/\lambda$.

The cdf is
\begin{equation}F(x; \mu, \lambda)=\Phi(z_1)+\exp(2\lambda/\mu)\Phi(z_2),\label{eq:igf}\end{equation}
where $z_1=(\lambda/x)^{1/2}(x/\mu-1)$, $z_2=-(\lambda/x)^{1/2}(x/\mu+1) $ and  $\Phi$ is the normal cdf.
The distribution has the additive property that the sum of two IG random variables is IG if $\mu^2/\lambda$ is the same for both variables.
Hence the cdf $F^{(n)}$ of the sum of $n$ iid. r.v.s is given by (\ref{eq:igf}) with $\mu\rightarrow n\mu, \lambda\rightarrow n^2\lambda$.

The RP-IG$(\mu,\lambda)$ discrete distribution is then computed from this as in the main text, and for the ERP-IG distribution $I_n$ in (\ref{eq:in0}), now called $K_n$,
is evaluated using integration by parts. Exploiting the fact that if $X$ is IG$(\mu,\lambda)$, then $Y=\mu^2/X$ has the length-biased pdf $yf(y)/\mu$
and using the Gaussian symmetry property $\Phi(-x)=1-\Phi(x)$ yields
\[K_n(t)=(t-\mu)\Phi(z_1)+(t+\mu)\exp(2n\lambda/\mu)\Phi(z_2),\]
where  in $z_1, z_2$ we have $\mu\rightarrow n\mu, \lambda\rightarrow n^2\lambda$.
The probabilities $Q_n$ of the ERP-IG$(\lambda,\mu)$ distribution can now be computed.

The moments are as before, with asymptotic variance now $t/\lambda+(1/6-(1/2)(\mu/\lambda)^2)$.

To generate random numbers from the IG distribution, the method quoted in \citet{gentle} or \cite{chik} is efficient and simple to program. Thus random numbers from the RP-IG distribution can be found as 
for the ERP-$\gamma$ distribution.
For the ERP-IG distribution, one can generate the first number from the length-biased distribution as before. Using the connection between the length-biased IG and IG distributions
already mentioned gives the length-biased r.v. $Y=\mu^2/X$, where $X \sim \text{IG}(\mu, \lambda)$. This is then multiplied by $U$, a uniformly-distributed random variable, as done for the ERP-$\gamma$ distribution.

To regress $\lambda, \mu$ on covariates, one can reflect that $\phi=\lambda/\mu$ plays the \role~of a shape parameter (the coefficient of variation is $\phi^{-1/2}$)
and so is analogous to $\beta$ for the gamma distribution, and can be kept constant, while $\lambda/\mu^2$, the hazard function in the exponential tail, is analogous to $\alpha$, and should depend on the covariates.
The same conclusion follows by equating the formulae for means and variances between the gamma and IG distributions.
Hence we take $\mu \propto \exp(-{\boldsymbol \beta}^T{\bf x})$, $\lambda \propto \exp(-{\boldsymbol \beta}^T{\bf x})$.

The fits to the number of births dataset was worse than for the Poisson distribution. The IG distribution with coefficient of dispersion equal to unity differs from the Poisson
in having a higher probability of zero, a peak at higher counts, and a shorter tail. The RP-IG and ERP-IG distributions can be over or underdispersed. For large $t$, the criterion for overdispersion is that $\lambda < \mu$,
but for small $t$ a value of $\lambda$ of somewhat less will give overdispersion.

The computational conclusion is that the ERP-IG distribution is even easier to compute with than the ERP-$\gamma$, requiring only the ubiquitous error function.
Further, generation of gamma-distributed random variables is not easy, although most platforms will have routines that can do this. The corresponding problem
for the IG distribution is trivial, requiring only the generation of Gaussian and uniform random numbers.

\clearpage

\section*{Figures and tables}

\begin{table}[h]
\begin{tabular}{|l|c|c|c|c|c|} \hline
Model & $\alpha$ & $\beta$ & param & param & $-\ell$ \\ \hline
Poisson & 2.38 & -& -& -& 2186.8 \\ \hline
RP-$\gamma$ & 2.86 & 1.16 & - & - & 2182.5 \\ \hline
ERP-$\gamma$ & 2.74 & 1.15& -& -&2181.9 \\ \hline
ERP-$\gamma$ $\beta$ mixture& 3.98&1.95&$\beta_2$ 0.93&$w$ 0.85&2137.6 \\ \hline
ERP-$\gamma$ $\alpha$ mixture&10.25&1.81&$\alpha_2$ 3.83& $w$ 0.077 &2138.1 \\ \hline
RP-$\gamma$(3) & 2.38 & 0.87 & $\delta$ 0.66& - & 2132.6 \\ \hline
\end{tabular}
\caption{\label{tab1}Fits of 6 models to the fertility data}
\end{table}
\begin{table}[h]
\begin{tabular}{|l|l|l|l|l|} \hline
&RP-$\gamma$ &&ERP-$\gamma$ & \\ \hline
Variable & Coeff. & se & Coeff. & se \\ \hline
$\hat\alpha$& 4.74& 1.20& 4.36&1.11\\ \hline
$\hat\beta$& 1.44& 0.071&1.39&0.063 \\ \hline
$-\ell$ &2078.22 &-& 2076.92&-\\ \hline
German&-.190&      0.059&  -.20&      0.062 \\ \hline    
Yrs schooling&0.032 & 0.026&   0.034&  0.027 \\ \hline   
Voc training&-.14&      0.036&  -.15 &      0.038 \\ \hline 
University&-.15 &     0.13&   -.16&      0.137 \\ \hline   
Catholic&0.21   &   0.058   &0.22 &      0.0614 \\ \hline  
Protestant&0.11 &     0.062& 0.11 &      0.066 \\ \hline       
Muslim&0.52   &   0.070&  0.55&      0.073 \\ \hline 
Rural&0.055&0.031   &0.059&  0.033 \\ \hline  
Year of birth&0.0023&  0.0019&   0.0026&  0.0020 \\ \hline 
Age at marriage&-.029&  0.0053&   -.031&  0.0057 \\ \hline    
\end{tabular}
\caption{\label{tab2}Fits of RP-$\gamma$ and ERP-$\gamma$ models to the fertility data with covariate regression.}
\end{table}
\end{document}